\documentclass[showpacs,preprintnumbers,twocolumn,
amsmath,amssymb,groupedaddress,superscriptaddress]{revtex4}
\usepackage{graphicx}
\usepackage{dcolumn}
\usepackage{bm}

\begin{document}

%\preprint{}

\title{Superscaling and Neutral Current Quasielastic Neutrino-Nucleus Scattering beyond the Relativistic Fermi Gas Model}

\author{A.N. Antonov}
\affiliation{Institute for Nuclear Research and Nuclear Energy, Bulgarian Academy of Sciences, Sofia 1784, Bulgaria}

\author{M.V. Ivanov}
\affiliation{Institute for Nuclear Research and Nuclear Energy,
Bulgarian Academy of Sciences, Sofia 1784, Bulgaria}

\author{M.B. Barbaro}
\affiliation{Dipartimento di Fisica Teorica, Universit\`a di Torino
and INFN, Sezione di Torino, Via P. Giuria 1, I-10125 Torino, Italy}

\author{J.A. Caballero}
\affiliation{Departamento de F\'\i sica At\'omica, Molecular y
Nuclear, Universidad de Sevilla, Apdo. 1065, 41080 Sevilla, Spain }

\author{E. Moya de Guerra}
\affiliation{Instituto de Estructura de la Materia, CSIC, Serrano
123, E-28006 Madrid, Spain}

\affiliation{Departamento de Fisica Atomica, Molecular y Nuclear,
Facultad de Ciencias Fisicas, Universidad Complutense de Madrid,
E-28040 Madrid, Spain}

\author{M.K. Gaidarov}
\affiliation{Institute for Nuclear Research and Nuclear Energy,
Bulgarian Academy of Sciences, Sofia 1784, Bulgaria}

\affiliation{Instituto de Estructura de la Materia, CSIC, Serrano
123, E-28006 Madrid, Spain}

%\date{\today}

\begin{abstract}

The superscaling analysis is extended to include quasielastic (QE)
scattering via the weak neutral current (NC) of neutrinos and
antineutrinos from nuclei. The scaling function obtained within the
coherent density fluctuation model (CDFM) (used previously in
calculations of QE inclusive electron and charge-changing (CC)
neutrino scattering) is applied to neutral current neutrino and
antineutrino scattering with energies of 1~GeV from $^{12}$C with a
proton and neutron knockout ($u$-channel inclusive processes). The
results are compared with those obtained using the scaling function
from the relativistic Fermi gas model and the scaling function as
determined from the superscaling analysis (SuSA) of QE electron
scattering.

\end{abstract}


\maketitle

\section[]{INTRODUCTION\label{sect1ant}}

The studies of the vast amount of inclusive electron scattering
world data have shown the existence of
$y$-scaling~\cite{ant01,ant02,ant03,ant04,ant05,ant06,ant07,ant08,ant09,ant10}
and superscaling (based on $\psi^\prime$-scaling variable) (see,
\emph{e.g.}~\cite{ant10,ant11,ant12,ant13,ant14,ant15,ant16,ant17,ant18,ant19,ant19a,ant20})
phenomena. A very weak dependence of the reduced cross section on
the momentum transfer $q$ (scaling of the first kind) has been
observed at excitation energies below the quasielastic peak for
large enough $q$. Scaling of the second kind (\emph{i.e.}~no
dependence of the reduced cross section on the mass number) has been
found to be excellent in the same region. When both types of scaling
occur one says that the reduced cross sections exhibit superscaling.
It has been shown (\emph{e.g.},
in~\cite{ant16,ant17,ant18,ant19,ant19a}) that the superscaling
phenomenon is related to the specific high-momentum tail of the
nucleon momentum distribution $n(k)$ at momenta $k > 2$~fm$^{-1}$
which is similar for all nuclei and is due to the short-range and
tensor correlations in the nuclear medium.

It has been observed also that above the QE peak the scaling of the
second kind is good, but scaling of the first kind is violated. The
latter occurs due to the excitation of a nucleon in the nucleus to a
delta-resonance which subsequently decays into a nucleon and a pion
(\emph{e.g.}, \cite{ant15,ant21}). Additionally, meson exchange
currents are known to violate the scaling
behavior~\cite{ant22,ant23,ant24,ant25}, although their effects
appear not to be the dominant ones~\cite{ant26}.

In~\cite{ant10,ant11} the theoretical concept of superscaling has
been introduced within the relativistic Fermi gas (RFG) model. As
pointed out in~\cite{ant13}, however, the actual dynamical physical
reason of the superscaling is more complex than that provided by the
RFG model. For instance, the QE scaling function in the RFG model is
$f_\text{RFG}^\text{QE}(\psi ') = 0$ for $\psi ' \leq -1$, whereas
the experimental scaling function $f^\text{QE}(\psi ')$ extends to
large negative values of $\psi '$ up to $\psi ' \approx -2$ in the
data for $(e,e')$ processes. Thus, the necessity to consider the
superscaling in theoretical methods which go beyond the RFG model
has arisen. One of them is the coherent density fluctuation model
(\emph{e.g.},~\cite{ant27,ant28}) being a natural extension of the
Fermi gas case to realistic finite nuclear systems. As pointed out
in~\cite{ant16,ant17,ant18,ant19}, in the CDFM both basic
quantities, density and momentum distributions are responsible for
the scaling and superscaling behavior in nuclei. The QE scaling
function in the CDFM $f(\psi ')$ agrees with the available
experimental data for $\psi '< 0$, including $\psi '\lesssim -1$.

In~\cite{ant29} the superscaling analyses of the electron scattering
for energies of several hundred MeV to a few GeV have been extended
to include not only QE processes but also those in which $\Delta
$-excitation dominates. Both QE- and $\Delta$-region scaling
functions $f^\text{QE}(\psi^\prime)$ and
$f^{\Delta}(\psi^\prime_\Delta )$ have been deduced in~\cite{ant29}
from phenomenological fits to the data for electron-nuclei
scattering cross sections. Generally, the theoretical microscopical
construction of the scaling function should take into account
final-state interactions (FSI). By using a relativistic mean field
for the final states, in~\cite{ant30,ant31} a scaling function with
asymmetric shape has been obtained being in agreement with the
experimental scaling function. Also an asymmetrical scaling function
in accordance with data has been obtained recently~\cite{ant33a}
within a semi-relativistic approach, based on improved
non-relativistic expansions, but with FSI described with the Dirac
equation-based potential. The asymmetry of the QE scaling function
in the CDFM has been introduced in a phenomenological
way~\cite{ant19} accounting for the role of FSI.

The analyses of the superscaling phenomenon and the present
knowledge of inclusive electron scattering off nuclei have induced
studies of neutrino scattering from nuclei on the same basis. This
makes it possible to explore fundamental questions of neutrino
reactions and neutrino oscillations in relation to hypothesis of
nonzero neutrino masses~\cite{ant32}. In~\cite{ant29} (see
also~\cite{ant30,ant33}) the scaling ideas have been inverted: given
the scaling functions one can just multiply by the elementary
charge-changing neutrino cross sections to obtain corresponding CC
neutrino and antineutrino cross sections on nuclei for intermediate
to high energies in the same region of excitation. In~\cite{ant26}
the scaling and superscaling ideas have been carried a step further
to include neutral current neutrino and antineutrino scattering
cross sections for scattering from $^{12}$C, namely for reactions
$^{12}$C($\nu,p)\nu$X, $^{12}$C(${\bar{\nu}},p){\bar{\nu}}$X
involving proton knockout and $^{12}$C($\nu,n)\nu$X,
$^{12}$C(${\bar{\nu}},n){\bar{\nu}}$X involving neutron knockout in
the QE regime. A number of other theoretical considerations
(\emph{e.g.},
\cite{ant34,ant35,ant36,ant37,ant38,ant39,ant40,ant41,ant42,ant43,ant44})
have been devoted to studies of both neutral- (\emph{e.g.},
\cite{ant34,ant35,ant36,ant37}) and charge-changing (\emph{e.g.},
\cite{ant35,ant36,ant37,ant38,ant39,ant40,ant41,ant42,ant43,ant44})
neutrino-nucleus scattering.

In~\cite{ant19} the QE- and $\Delta$-region scaling functions
obtained in the CDFM and within the modified parameter-free
theoretical approach~\cite{ant45} based on the light-front dynamics
method (LFD) (\emph{e.g.}, \cite{ant46,ant47}) have been applied to
describe the experimental data on differential cross sections of QE
inclusive electron scattering as well as to analyze charge-changing
neutrino scattering on the $^{12}$C nucleus for energies of the
incident particles from 1 to 2~GeV. It has been shown that the
results for electron scattering on $^{12}$C are close to those of
the superscaling analysis~\cite{ant15,ant29} and are quite different
from the RFG results, whereas the almost symmetric CDFM scaling
function leads to cross sections that are similar to the results of
the RFG model.

The aim of the present work is to extend the application of the CDFM
scaling function to calculations of neutral current neutrino and
antineutrino scattering cross sections from nuclei, \emph{e.g.} to
consider reactions on $^{12}$C as those in~\cite{ant26} and
mentioned above with proton and neutron knockout in the QE region.
We should note that, as it has been considered
in~\cite{ant26,ant48}, when one has an incident lepton, a scattering
with exchange of a $\gamma$, $W^\pm$ or $Z^0$ and the scattered
lepton (\emph{i.e.} a charged lepton) is detected, the $t$-channel
exchange of the corresponding boson is controlled. When, however,
the scattered lepton is a (not detected) neutrino or antineutrino,
and a knocked-out nucleon is detected, then the kinematics of the
$u$-channel are controlled. We also follow~\cite{ant26} on the
formalism for cross sections and, what is more crucial, in what
concerns the scaling ideas that interrelate the $t$- and $u$-channel
processes.

%Further in this work we follow the consideration from~\cite{ant26}
%of the $t$- and $u$-channel kinematics in the semi-leptonic
%electroweak processes, the cross section formalism and the ideas of
%scaling when interrelating $t$- and $u$-channel processes.

The paper is organized in the following way: the theoretical
scheme is given in Sec.~\ref{sect2ant}. It includes the formalism
for $u$-channel scattering including briefly the kinematics, cross
sections and scaling as well as the main relationships of the CDFM
used in the superscaling analysis. The results of NC neutrino and
antineutrino scattering cross sections on $^{12}$C are presented
and discussed in Sec.~\ref{sect3ant}. The conclusions are
summarized in Sec.~\ref{sect4ant}.

\section[]{THE THEORETICAL SCHEME\label{sect2ant}}

\subsection[]{Kinematics. Cross sections and scaling\label{sect2Aant}}

\begin{figure}[b]
\centering
\includegraphics[width=70mm]{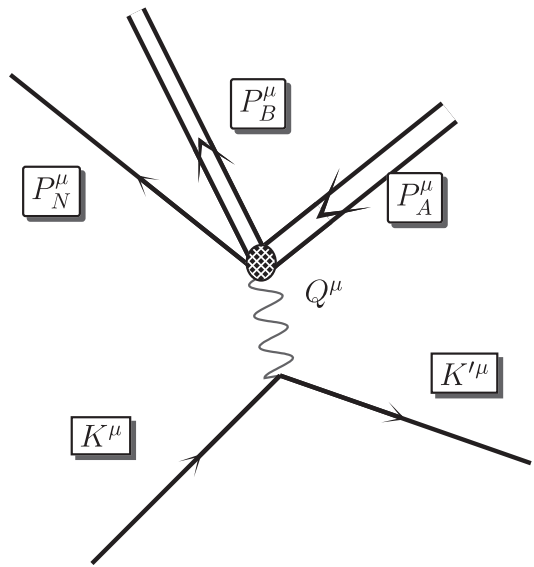}
\caption[]{The kinematics for semi-leptonic nucleon knockout
reactions in the one-boson-exchange approximation.\label{fig1ant}}
\end{figure}

We consider the semi-leptonic quasi-free scattering from nuclei in
Born approximation, assuming that the inclusive cross sections are
well represented by the sum of the integrated semi-inclusive proton
and neutron emission cross sections~\cite{ant26}. The kinematics for
semi-leptonic nucleon knockout reactions in the one-boson-exchange
approximation is presented in Fig.~\ref{fig1ant}.

A lepton with 4-momentum $K^{\mu }=(\epsilon ,\mathbf{k})$ scatters
to another lepton with 4-momentum $K^{\prime \mu }=(\epsilon
^{\prime },\mathbf{k}^{\prime })$, exchanging a vector boson with
4-momentum $Q^{\mu }=K^{\mu }-K^{\prime \mu }$. The lepton energies
are $\epsilon =\sqrt{m^{2}+k^{2}}$ and $\epsilon ^{\prime }
=\sqrt{m^{\prime 2}+k^{\prime 2}}$, where the masses of the initial
and final lepton  $m$ and $(m^{\prime })$ are assumed to be equal to
zero for NC neutrino scattering. In the laboratory system the
initial nucleus being in its ground state has a 4-momentum
$P_{A}^{\mu } =(M_{A}^{0},0)$, while the final hadronic state
corresponds to a proton or neutron with 4-momentum
$P_{N=p~\text{or}~n}^{\mu }=(E_{N},{{\bf p}}_N)$ and an unobserved
residual nucleus with 4-momentum $P_{B}^{\mu
}=(E_{B},\mathbf{p}_{B})$. Usually the missing momentum $\mathbf{p}
\equiv -\mathbf{p}_{B}$ and the excitation energy $\mathcal{E}\equiv
E_{B} - E_{B}^{0}$, with $E_{B}^{0} =\sqrt{\left( M_{B}^{0}\right)
^{2}+p^{2}}$ are introduced, $M_{B}^{0}$ being the ground-state mass
of the daughter nucleus. We assume for NC neutrino scattering that
the neutrino beam momentum is specified and the outgoing proton (or
neutron) is detected (for details see~\cite{ant12,ant13,ant26}). The
exchanged 4-momentum in the $u$-channel is defined as
\begin{equation}
Q^{\prime \mu } \equiv K^{\mu }-P_{N}^{\mu } =(\omega ^{\prime
},\mathbf{q}^{\prime }).  \label{new1}
\end{equation}
%with
%\begin{equation}
%q^{\prime }=|\mathbf{q}^{\prime }|=\sqrt{k^{2}+p_{N}^{2}-2kp_{N}\cos
%\theta _{kp_{N}}} .  \label{new2}
%\end{equation}
Details on the kinematics and integration limits involved in NC
neutrino-nucleus scattering are given in~\cite{ant26}.
%
%One can use a coordinate system with the $z$-axis along
%$\mathbf{q}^{\prime }$, with $\mathbf{k}$ and $\mathbf{p}_{N}$ lying
%in the $xz$-plane. Then the vectors $\mathbf{k}^{\prime }$ and
%$\mathbf{p}=\mathbf{k}^{\prime }-\mathbf{q}^{\prime }$ lie in a
%plane forming an angle $\phi^{\prime }$ with the $xz$-plane. The six
%kinematic variables which determine exclusive process in
%Fig.~\ref{fig1ant} can be chosen to be ($k$, $p_{N}$, $\theta
%_{kp_{N}}$, $p$, $\mathcal{E}$, $\phi^{\prime }$). The $u$-channel
%inclusive cross section for fixed $(k,p_{N},\theta _{kp_{N}})$ can
%be obtained by integrating over the allowed region in the ($p$,
%${\mathcal{E}}$) plane and over the azimuthal angle $\phi ^{\prime
%}$ ($0\leq \phi ^{\prime }\leq 2\pi $). Details on the integrations
%are given in~\cite{ant26}.

The usual procedure for calculating the $(l,l' N)$ cross section
includes the Plane Wave Impulse Approximation (PWIA) and
integrations over all unconstrained kinematic variables. It is shown
in~\cite{ant26} that the inclusive cross section in the $u$-channel
can be written after some approximations in the following form:
\begin{equation}\label{new3}
\frac{d\sigma}{d\Omega_{N} dp_N} \simeq \overline{\sigma}_{sn}^{(u)}
F(y',q') ,
\end{equation}
where
\begin{equation}\label{new4}
F(y',q') \equiv \int\limits_{{\cal D}_u} p dp \int \frac{d{\cal
E}}{E} \Sigma \simeq F(y'),
\end{equation}
provided the effective NC single nucleon (s.n.) cross section
\begin{multline}\label{new5}
\overline{\sigma}_\text{s.n.}^{(u)}= \frac{1}{32\pi
\epsilon}\frac{1}{q^\prime} \left(\frac{p_N^2}{E_N}\right) g^4
\int\limits_0^{2\pi} \frac{d\phi'}{2\pi}l_{\mu\nu}(\mathbf{k},\mathbf{k}')\\
\times w^{\mu\nu}({\mathbf{p}},{\mathbf{p}}_N) D_V(Q^2)^2
\end{multline}
is almost independent of $(p,{\cal E})$ for constant
$(k,p_N,\theta_{kp_N})$. In Eq.~(\ref{new5}) $l_{\mu\nu}$ and
$w^{\mu\nu}$ are the leptonic and s.n. hadronic tensor,
respectively, and $D_V(Q^2) $ is the vector boson
propagator~\cite{ant26}. In Eq.~(\ref{new3}) $y'$ is the scaling
variable naturally arising in the $u$-scattering kinematics,
analogous to the usual $y$-scaling variable for $t$-scattering. The
scaling function $F(y')$ obtained within a given approach can be
used to predict realistic NC cross sections. Assuming that the
domains of integration ${\cal D}_u$ (in the $u$-channel) and ${\cal
D}_t$ (in the $t$-channel) are the same or very similar, the results
for the scaling function obtained in the case of inclusive electron
scattering (where ${\cal D}_t$ works) can be used in the case of NC
neutrino reactions. It is pointed out in~\cite{ant26} that ${\cal
D}_t$ and ${\cal D}_u$ differ significantly only at large ${\cal E}$
(also at large $p$, but there one believes that the semi-inclusive
cross sections are negligible). So, given that the semi-inclusive
cross sections are dominated by their behavior at low ${\cal E}$ and
low $p$, one expects the results of the integrations in the $t$- and
$u$-channel to be very similar, and thus the scaling functions will
be essentially the same in both cases.

The RFG $u$-channel $\psi$-variable is introduced in the
form~\cite{ant26}:
\begin{equation}\label{new6}
\psi^{(u)}_{RFG}=s\sqrt{\frac{m_N}{T_F}}
\left[\sqrt{1+\left(\frac{y^{(u)}_\text{RFG}}{m_N}\right)^2}-1\right]^{1/2},
\end{equation}
where
\begin{equation}\label{new7}
y^{(u)}_{RFG}= s \frac{m_N}{\tau'} \left[
\lambda'\sqrt{\tau'^2\rho'^2+\tau'}-\kappa'\tau'\rho'\right]
\end{equation}
is the RFG $y$-scaling variable for the $u$-channel and corresponds
to the minimum momentum required for a nucleon to participate in the
NC neutrino-nucleus scattering. The dimensionless kinematic
quantities in Eq.~(\ref{new7}) are given by: $\kappa'\equiv q'/2
m_N$, $\lambda'\equiv \omega'/2 m_N$, $\tau'=\kappa'^2-\lambda'^2$
and defined $\rho'\equiv 1-\dfrac{1}{4\tau'} (1-m'^2/m_N^2)$. The
sign $s$ is
\begin{equation}\label{new8}
s\equiv{\rm sgn}\left\{ \frac{1}{\tau'} \left[
\lambda'\sqrt{\tau'^2\rho'^2+\tau'}-\kappa'\tau'\rho'\right]\right\}
 .
\end{equation}

The physical meaning of $\psi^{(u)}_\text{RFG}$ is the minimum
kinetic energy of the nucleon participating in the reaction. The RFG
scaling function is found to be:
\begin{equation}\label{new9}
 F_\text{RFG}(\psi^{(u)}_\text{RFG}) = \frac{3}{4}
k_F\left(1-\psi^{(u)2}_\text{RFG}\right)
\Theta\left(1-\psi^{(u)2}_\text{RFG}\right).
\end{equation}
As noted in~\cite{ant26}, if the s.n.~cross section is smoothly
varying within the $(p,{\cal E})$ integration region, the
differential cross section in the RFG can be factorized as shown in
Eq.~(\ref{new3}) with the scaling function from Eq.~(\ref{new9}). In
this work, however, we use the scaling function calculated in the
CDFM model which is beyond the RFG model (see
subsection~\ref{sect2Bant}).

The basic relationships used to calculate the s.n.~cross sections
are given in~\cite{ant26}. This concerns the leptonic and hadronic
tensors and the response and structure functions. The H\"{o}hler
parametrization for the single-nucleon form factors~\cite{ant50} is
used, ignoring the strangeness content of the nucleon.

\subsection[]{QE scaling function in the CDFM\label{sect2Bant}}

In this subsection we present briefly the main expressions
concerning the QE scaling function $f^{\text{QE}}(\psi^\prime)$
within the CDFM~\cite{ant27,ant28} (which is a natural extension of
the RFG model).
%$\psi^\prime$ has the form:
%\begin{equation}\label{eq28'ant}
%\psi^\prime\equiv \dfrac{1}{\sqrt{\xi_F}}
%\dfrac{\lambda^\prime-\tau^\prime}{\sqrt{(1+\lambda^\prime)\tau^\prime+
%\kappa^\prime\sqrt{\tau^\prime(1+\tau^\prime)}}}
%\end{equation}with $\eta_F \equiv k_F/m_N$ and
%$\xi_F\equiv \sqrt{1+\eta_F^2}-1$. The scaling function
This function was obtained (see~\cite{ant19} and references therein)
in two ways that were shown to be equivalent: on the basis of the
local density distribution ($\rho(r)$) and on the basis of the
nucleon momentum distribution ($n(k)$). Generally, the total CDFM
scaling function is expressed by the sum of the proton
$f_{p}^\text{QE}(\psi^{\prime})$ and neutron
$f_{n}^\text{QE}(\psi^{\prime})$ scaling functions, which are
determined by the proton and neutron densities $\rho_{p}(r)$ and
$\rho_{n}(r)$ (or by corresponding momentum distributions),
respectively:
\begin{equation}
f^{QE}(\psi^{\prime})=\dfrac{1}{A}[Zf_p^\text{QE}(\psi^{\prime})+Nf_n^\text{QE}(\psi^{\prime})].\label{new10}
\end{equation}

The CDFM scaling function gives a good description of the
superscaling phenomenon. In the consideration
in~\cite{ant16,ant17,ant18} it has a symmetric form for negative and
positive values of $\psi^\prime$. The maximum value of
$f^\text{QE}(\psi^{\prime})$ in CDFM (and in RFG) is 3/4 whereas,
however, the empirical ``universal'' scaling function extracted
in~\cite{ant29} reaches 0.6 and has a markedly asymmetric shape.
Also, an asymmetric shape of $f(\psi^\prime)$ has been found
in~\cite{ant30,ant31} from calculations for ($e,e^\prime$) and
($\nu,\mu$) reactions based on the relativistic impulse
approximation with FSI using the relativistic mean-field potential.

In~\cite{ant19} we limited our CDFM approach to phenomenology when
considering the asymmetric shape and the maximum value of the QE
$f(\psi^\prime)$. The role of all the effects that lead to asymmetry
has been simulated by imposing asymmetry on the RFG scaling function
(and, correspondingly, on the CDFM one) by introducing a parameter
which gives the correct maximum value of the scaling function
($c_{1}$ in our notations below) and also an asymmetric tail in
$f^\text{QE}(\psi^{\prime})$ for $\psi^{\prime}\geq 0$. The proton
and neutron scaling functions in Eq.~(\ref{new10}) are presented as
sums of scaling functions for negative
($f_{p(n),1}^\text{QE}(\psi^{\prime})$) and positive
($f_{p(n),2}^\text{QE}(\psi^{\prime})$) values of $\psi^\prime$:
\begin{equation}
f_{p(n)}^\text{QE}(\psi^{\prime})=
f_{p(n),1}^\text{QE}(\psi^{\prime})+f_{p(n),2}^\text{QE}(\psi^{\prime}).
\label{new11}
\end{equation}
In Eq.~(\ref{new11})
\begin{multline}
f_{p(n),1}^\text{QE}(\psi^{\prime})=\!\!\!\!\int\limits_{0}^{\alpha_{p(n)}/(k^{p(n)}_{F}
|\psi^{\prime}|)}\!\!\!\!dR
|F_{p(n)}(R)|^{2}f_\text{RFG,1}^{p(n)}(\psi'(R)),\\
\psi^\prime\leq0, \label{new12}
\end{multline}
\begin{multline}
f_{p(n),2}^\text{QE}(\psi^{\prime})=\!\!\!\!
\int\limits_{0}^{c_{2}\alpha_{p(n)}/(k_{F}^{p(n)}\psi^{\prime})}\!\!\!\!\!\!
dR
|F_{p(n)}(R)|^{2} f_\text{RFG,2}^{p(n)}(\psi'(R)),\\
\psi^{\prime}\geq 0, \label{new14}
\end{multline}
where
\begin{equation}
f_\text{RFG,1}^{p(n)}(\psi^\prime(R)) = c_1\left[ 1-\left(
\frac{k^{p(n)}_FR|\psi^\prime|}{\alpha_{p(n)}}
\right)^{2}\right],~\psi^\prime\leq0 \label{new13}
\end{equation}
and
\begin{equation}
f_\text{RFG,2}^{p(n)}(\psi^\prime(R)) = c_1\exp\left[
-\frac{k_{F}^{p(n)}R\psi^\prime}{c_2\alpha_{p(n)}}
\right],~\psi^\prime\geq0. \label{new15}
\end{equation}

In Eqs.~(\ref{new12}) and (\ref{new14}) the proton and neutron
weight functions are obtained from the corresponding proton and
neutron densities
\begin{gather}
\left|F_{p(n)}(R)\right|^2=-\dfrac{4\pi
R^3}{3Z(N)}\left.\dfrac{d\rho_{p(n)}(r)}{dr}\right|_{r=R},
\label{new16}\\
\alpha_{p(n)}=\left[\dfrac{9\pi Z(N)}{4}\right]^{1/3},
\label{new17}\\
\int\limits_{0}^{\infty}\rho_{p(n)}(\mathbf{r})d\mathbf{r}=Z(N)
\label{new18}
\end{gather}
and the Fermi-momentum for the protons and neutrons can be
calculated using the expression
\begin{equation}
k_{F}^{p(n)}=\alpha_{p(n)}\int\limits_{0}^{\infty}dR
\frac{1}{R}|{F}_{p(n)}(R)|^{2}. \label{new19}
\end{equation}
The functions are normalized as follows:
\begin{gather}
\int\limits_{0}^{\infty}|F_{p(n)}(R)|^{2}dR=1, \label{new20}\\
\int\limits_{-\infty}^{\infty}f_{p(n)}^\text{QE}(\psi^{\prime})d\psi^{\prime}=1.
\label{new21}
\end{gather}

From the normalization of the total QE scaling function
\begin{equation}
\int\limits_{-\infty}^{\infty}f^\text{QE}(\psi^{\prime})d\psi^{\prime}=1\label{new22}
\end{equation}
one can get relationship between $c_2$ and $c_1$:
\begin{equation}c_2=\dfrac{1-\dfrac{2}{3}c_1}{c_1\left(\dfrac{e-1}{e}\right)}\simeq
\dfrac{1-\dfrac{2}{3}c_1}{0.632c_1}.\label{new23}\end{equation} The
value of $c_2=1$ corresponds to $c_1=\dfrac{3}{4}$. The asymmetry of
the scaling function increases with the decrease of $c_1$  from
$\dfrac{3}{4}$.

In~\cite{ant19} also a parabolic form of
$f_\text{RFG,2}^{p(n)}(\psi^\prime(R))$
\begin{equation}
f_\text{RFG,2}^{p(n)}(\psi^\prime(R)) =
c_{1}\left[1-\left(\frac{k_{F}^{p(n)}R\psi^{\prime}}{c_2\alpha_{p(n)}}\right
)^{2}\right ],~\psi^\prime\geq0. \label{new24}
\end{equation}
instead of the exponential one in Eq.~(\ref{new15}) was considered.
In this case $c_2=\dfrac{3}{2c_1}-1$.

As already mentioned, the QE- and $\Delta $-scaling functions
obtained in the CDFM and in the LFD approach were applied
in~\cite{ant19} to describe experimental data on differential cross
sections of inclusive electron scattering by $^{12}$C at large
energies and transferred momenta as well as to calculate QE
charge-changing neutrino-nuclei reaction cross sections. It was
shown in the case of the electron scattering that the results
obtained when asymmetric scaling function $f^\text{QE}(\psi^\prime)$
($c_{1}^\text{QE}=0.63$) with
$f_\text{RFG,2}^{p(n)}(\psi^\prime(R))$ from Eq.~(\ref{new24}) is
used agree with the data in cases when the transferred momentum in
the position of the maximum of the QE peak extracted from data
($\omega^\text{QE}_\text{exp}$) is
$q^\text{QE}_\text{exp}<450~\text{MeV/c}\approx2k_F$ and
underestimate them when $q_{exp}^\text{QE}\geq 450$~MeV/c in the
region close to the QE peak. The almost symmetric scaling function
$f^\text{QE}(\psi^\prime)$ ($c_{1}^\text{QE}=0.72$) leads to results
in agreement with the data in the region of the QE peak in cases
when $q_{exp}^\text{QE}\geq 450$~MeV/c, whereas the data are
overestimated in cases where
$q^\text{QE}_\text{exp}<450~\text{MeV/c}$. As can be seen in
Section~\ref{sect3ant}, the use of the exponential form of
$f_\text{RFG,2}^{p(n)}(\psi^\prime(R))$ [Eq.~(\ref{new15})] instead
of the parabolic one [Eq.~(\ref{new24})] was imposed by the aim for
a better description of the experimental data for the quasielastic
scaling function $f^\text{QE}(\psi^\prime)$ for $\psi^\prime\geq0$
(see Fig.~\ref{fig6ant} for $f^\text{QE}(\psi^\prime)$ in the case
of $^{12}$C nucleus). In the cases of CC neutrino and antineutrino
$(\nu_{\mu},\mu^{-})$ and $(\bar{\nu}_{\mu},\mu^{+})$ reactions on
$^{12}$C for energies of the incident particles from 1 to 2~GeV the
results obtained by using the asymmetric CDFM scaling function
$f^\text{QE}(\psi^\prime)$ ($c_{1}^\text{QE}=0.63$) are close to
those of SuSA~\cite{ant15, ant29} and are different from the RFG
model results, whereas the almost symmetric CDFM scaling function
$f^\text{QE}(\psi^\prime)$ ($c_{1}^\text{QE}=0.72$) leads to cross
sections that are similar to the results of the RFG model.

\section[]{RESULTS OF CALCULATIONS AND DISCUSSION\label{sect3ant}}

\begin{figure*}
\centering
\includegraphics[width=137mm]{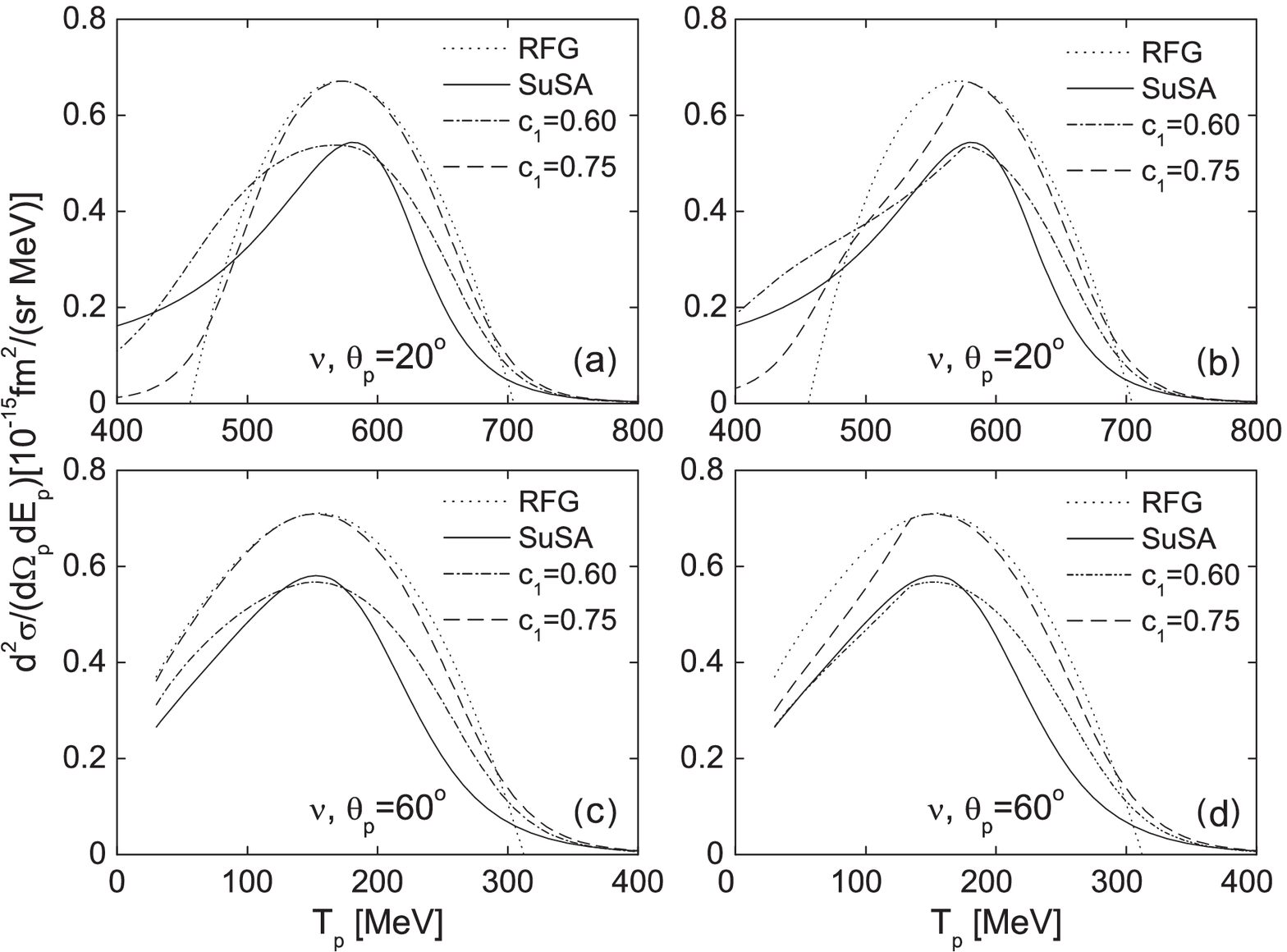}
\caption[]{Quasielastic differential cross section for neutral
current neutrino scattering at 1~GeV from $^{12}$C for proton
knockout at $\theta_p=20^\circ$ (a,b) and $60^\circ$ (c,d) using the
CDFM scaling function [Eqs.~(\ref{new10})--(\ref{new13}),
(\ref{new24}) for (a,c) and Eqs.~(\ref{new10})--(\ref{new15}) for
(b,d)] with $ c_1=0.60$ (dash-dotted line) and $c_1=0.75$ (dashed
line). The RFG results are given by dotted line and the results
using the empirical scaling function~\cite{ant26} are presented by
solid line (SuSA).\label{fig2ant}}

\bigskip

\centering
\includegraphics[width=137mm]{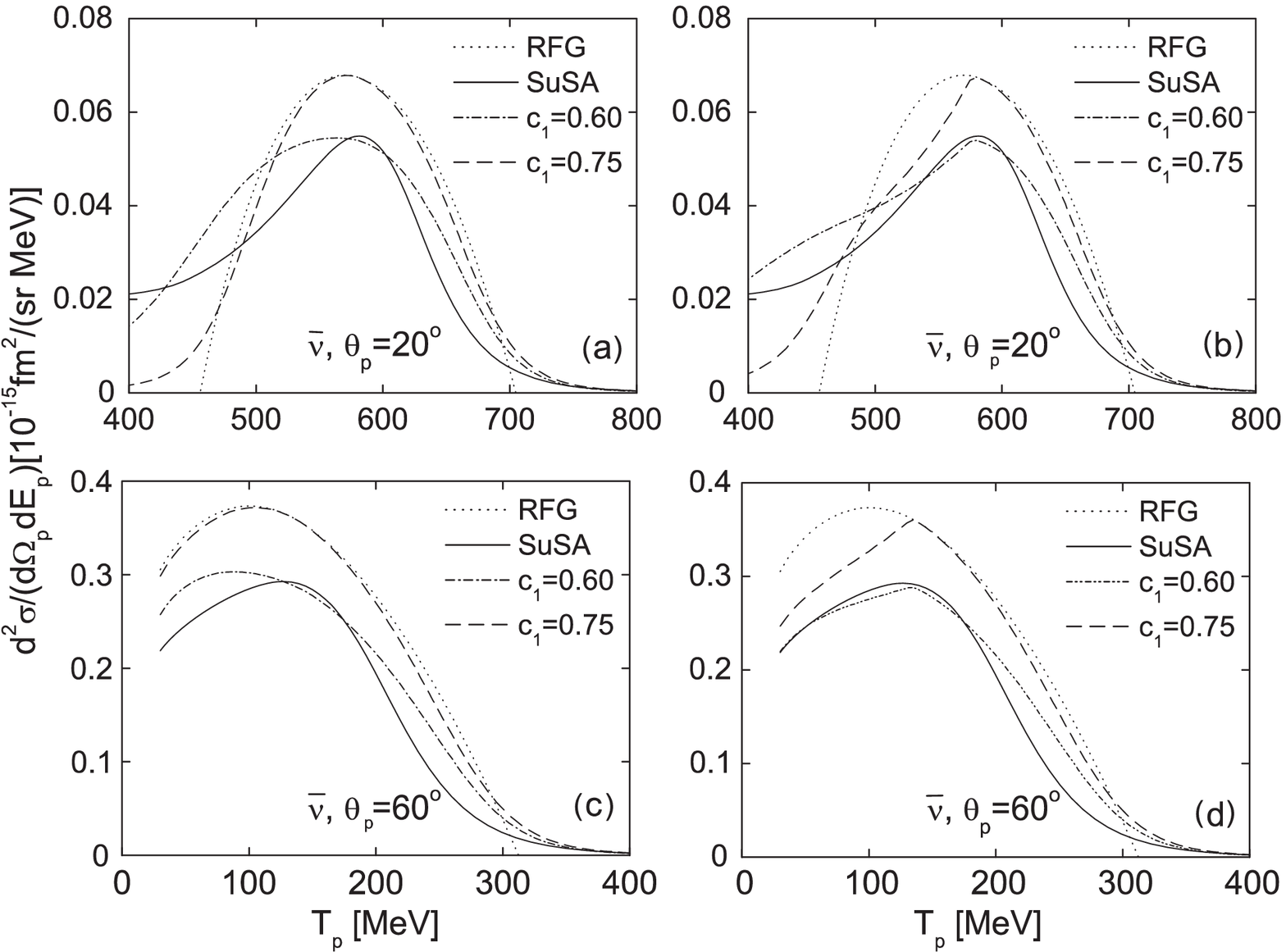}
\caption[]{The same as in Fig.~\ref{fig2ant} for neutral current
antineutrino scattering.\label{fig3ant}}
\end{figure*}

\begin{figure*}
\centering
\includegraphics[width=140mm]{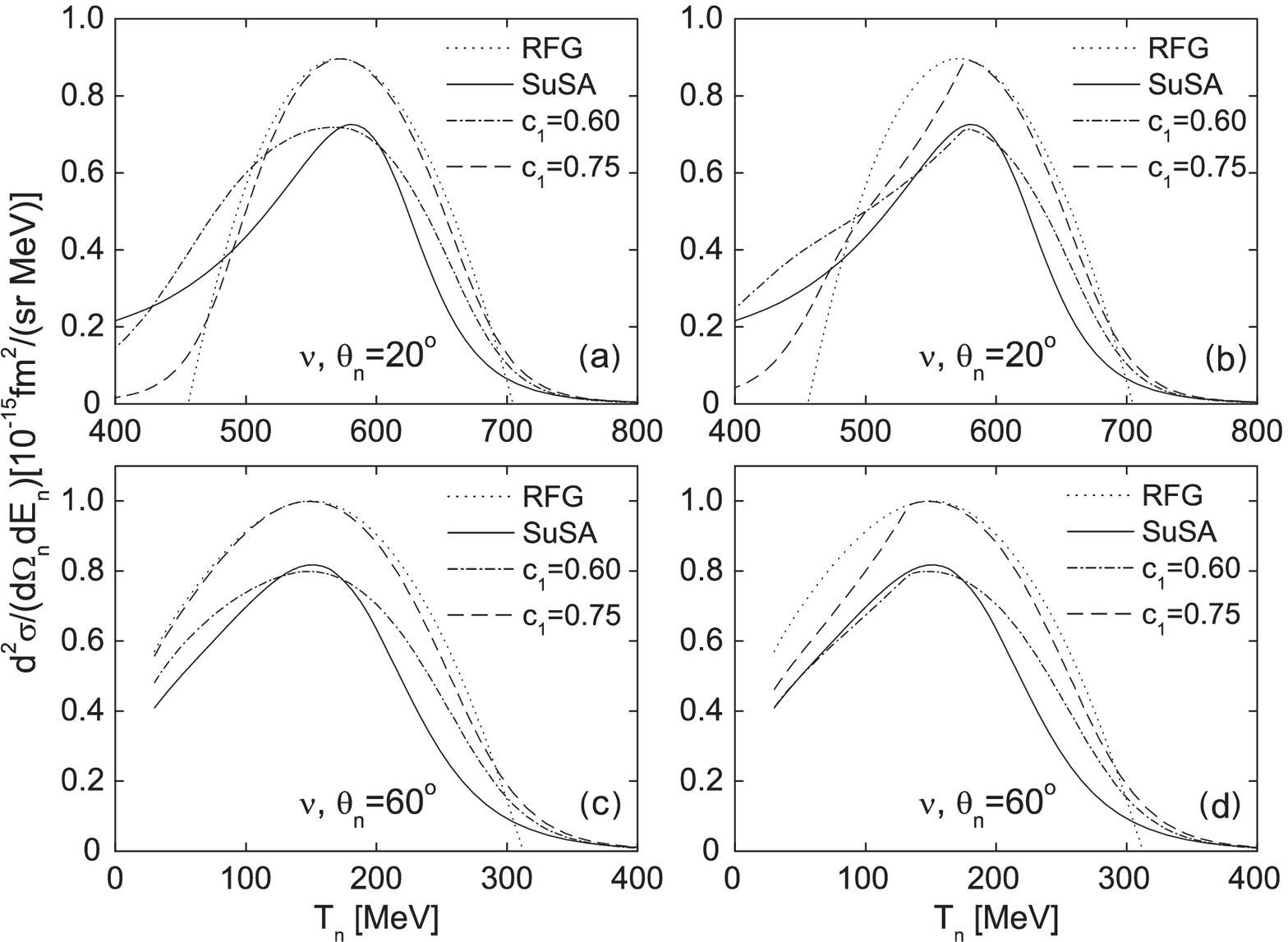}
\caption[]{The same as in Fig.~\ref{fig2ant} for neutral current
neutrino scattering showing the neutron knockout
case.\label{fig4ant}}

\bigskip

\centering
\includegraphics[width=140mm]{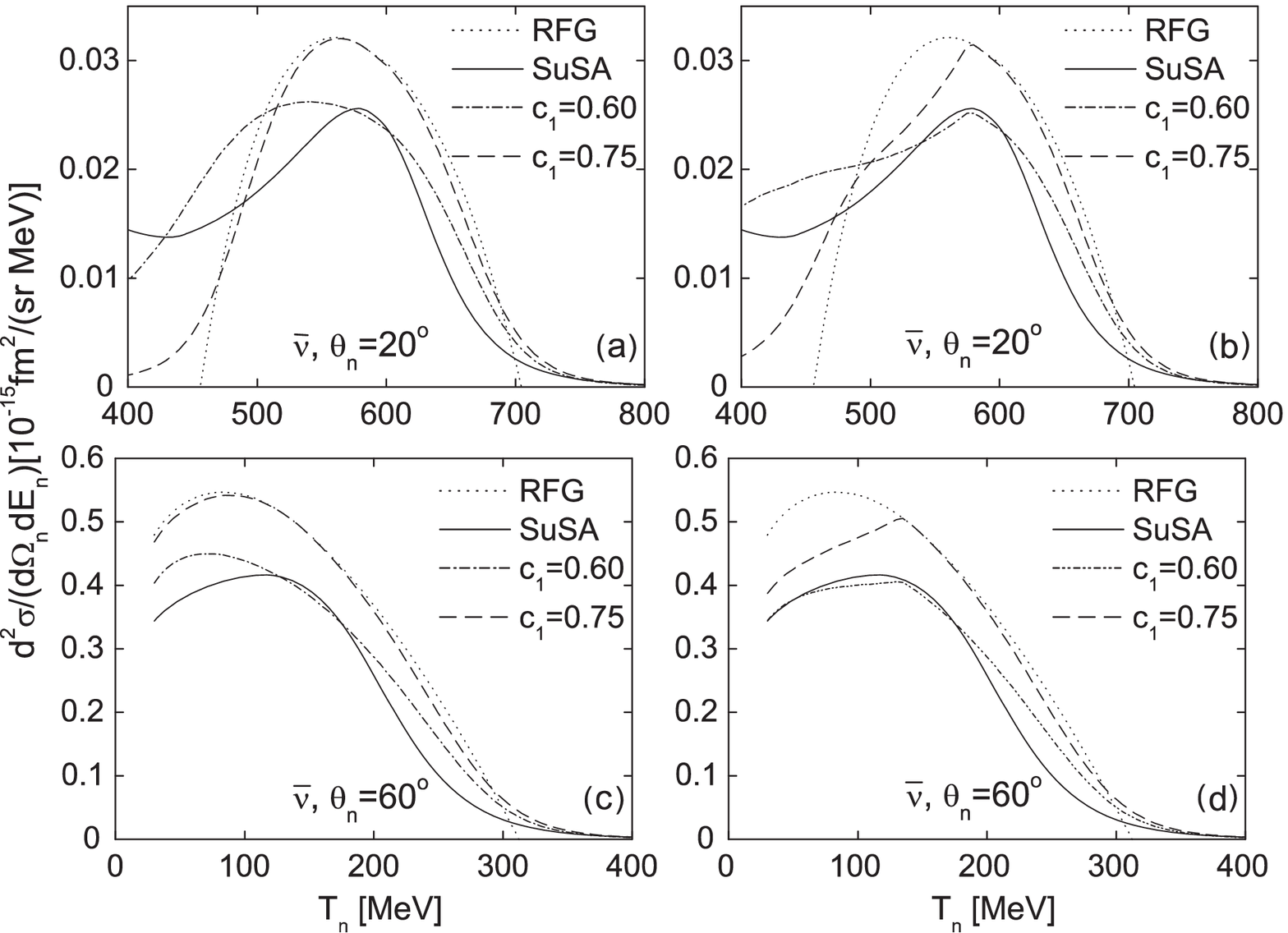}
\caption[]{The same as in Fig.~\ref{fig2ant} for neutral current
antineutrino scattering showing the neutron knockout
case.\label{fig5ant}}
\end{figure*}

\begin{figure*}
\centering
\includegraphics[width=150mm]{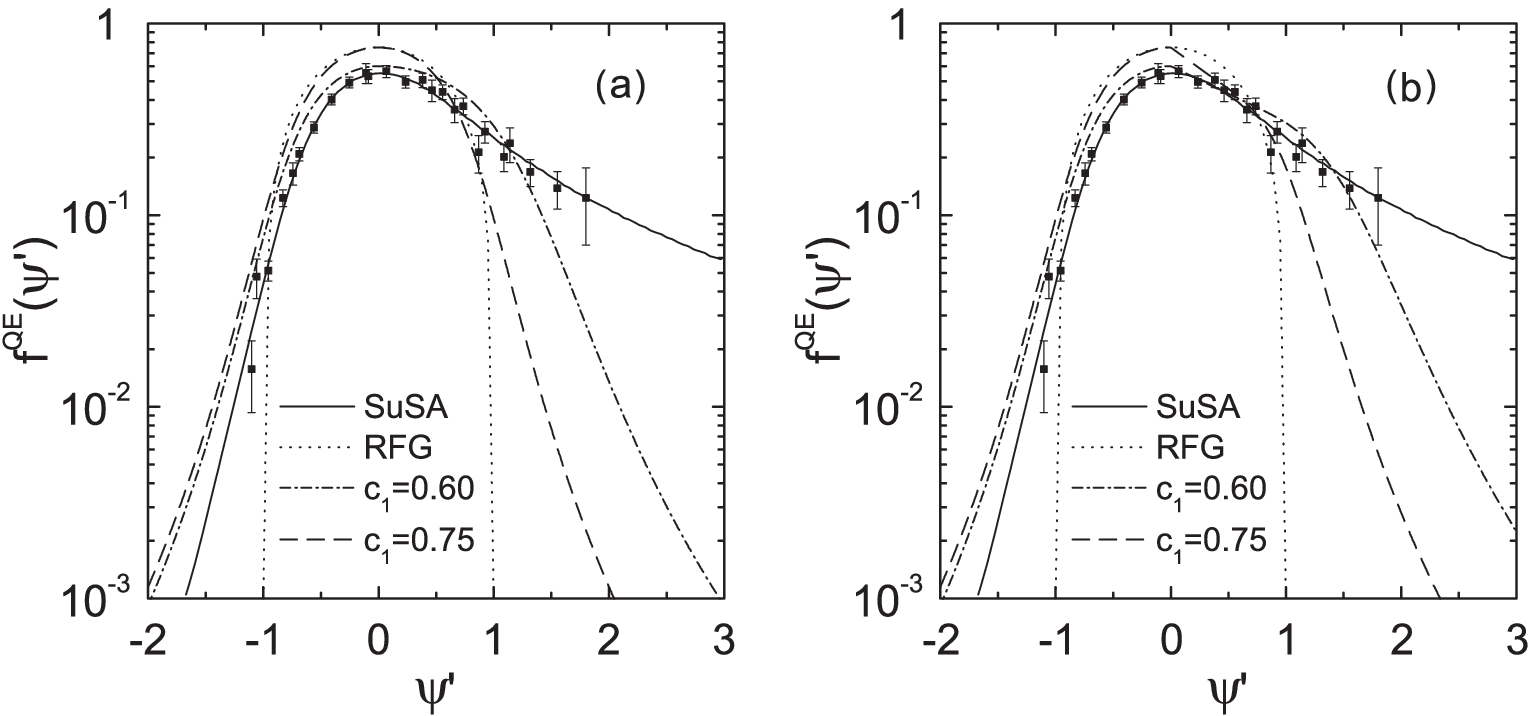}
\caption[]{(a) The quasielastic  scaling function
$f^\text{QE}(\psi^\prime)$ for $^{12}$C calculated in the CDFM using
Eqs.~(\ref{new10})--(\ref{new13}), (\ref{new24}) with $c_1=0.60$
(dash-dotted line) and $c_1=0.75$ (dashed line) in comparison with
the result of the RFG model (dotted line) and with the results from
the SuSA (solid line). The experimental data (black squares) are
taken from~\cite{ant29}; (b) the same as in (a) with the CDFM
scaling function calculated by using
Eqs.~(\ref{new10})--(\ref{new15}).\label{fig6ant}}
\end{figure*}

In this section we present in Figs.~\ref{fig2ant}--\ref{fig5ant} the
results of the calculations of the cross sections of neutral current
neutrino (Figs.~\ref{fig2ant} and~\ref{fig4ant}) and antineutrino
(Figs.~\ref{fig3ant} and~\ref{fig5ant}) scattering at 1~GeV from
$^{12}$C with a knockout of proton (Figs.~\ref{fig2ant}
and~\ref{fig3ant}) and neutron (Figs.~\ref{fig4ant}
and~\ref{fig5ant}) as a function of the kinetic energy of the
ejected nucleon. The calculations are performed for two values of
the proton or neutron angle, namely $20^\circ$ (a,b) and $60^\circ$
(c,d).

The two ingredients of the cross section, namely the s.n. cross
section and the QE scaling function, are calculated according to the
theoretical scheme presented in subsections~\ref{sect2Aant}
and~\ref{sect2Bant}, correspondingly. The NC neutrino and
antineutrino scattering cross sections are calculated using
Eqs.~(\ref{new3})--(\ref{new5}), while for the s.n. cross sections
we followed the consideration in~\cite{ant26}. The CDFM QE scaling
function for the $^{12}$C nucleus was calculated by means of
Eqs.~(\ref{new10})--(\ref{new24}) using in Eq.~(\ref{new16}) the
charge density of $^{12}$C and assuming that the proton and neutron
densities are the same. We used a symmetrized Fermi-type density
distribution~\cite{ant51} with the following values of the
half-radius $R_{1/2}$ and diffuseness $b$ parameters:
$R_{1/2}=2.470$~fm and $b=0.420$~fm. These parameter values lead to
charge rms radius equal to $2.47$~fm which coincides with the
experimental one~\cite{ant52}.

In Figs.~\ref{fig6ant}a and~\ref{fig6ant}b we present the
quasielastic CDFM scaling function $f^\text{QE}(\psi^\prime)$ for
$^{12}$C in comparison with the experimental data taken
from~\cite{ant29}, with the RFG result and with the SuSA result. The
results for the QE scaling function using the parabolic form of the
RFG scaling function for $\psi^\prime \geq0$,
[Eqs.~(\ref{new10})--(\ref{new13}),~(\ref{new24})] are given in
Fig.~\ref{fig6ant}a, while those obtained using the exponential form
of the RFG scaling function for $\psi^\prime \geq0$
[Eqs.~(\ref{new10})--(\ref{new15})] are given in
Fig.~\ref{fig6ant}b. The CDFM scaling function is given in
Figs.~\ref{fig6ant}a and~\ref{fig6ant}b for two values of the
parameter $c_1$: $c_1=0.75 $ and $0.60$. In the case of $c_1=0.75 $
$f^\text{QE}(\psi^\prime)$ is symmetric, while in the case with
$c_1=0.60 $ it is asymmetric. As can be seen in both cases the
scaling functions calculated using $c_1=0.60 $ are in better
agreement with the empirical data. This is true even in the interval
$\psi^\prime <-1$, whereas in the RFG model
$f_\text{RFG}(\psi^\prime)=0$ for $\psi^\prime \leq-1$.

The calculations of the quasielastic differential cross sections for
neutral current neutrino and antineutrino scattering in the proton
and neutron knockout cases whose results are presented in
Figs.~\ref{fig2ant}--\ref{fig5ant}, are performed by means of the
CDFM QE scaling function in both cases (using Eq.~(\ref{new15}) or
Eq.~(\ref{new24}) for $f_\text{RFG,2}^{p(n)}(\psi^\prime(R))$ in
Eqs.~(\ref{new10})--(\ref{new13})).

We would like to note the following features of the results.
Firstly, it can be seen that our results by using $c_1=0.60$
(\emph{i.e.} asymmetric $f^\text{QE}(\psi^\prime)$) are close to
those obtained in~\cite{ant26} from the SuSA showing a tail for
larger values of the kinetic energy $T_{p(n)}$ in contrast to the
RFG result. When using $c_1=0.75$ (\emph{i.e.} symmetric
$f^\text{QE}(\psi^\prime)$) our cross sections are close to those
from the RFG model. Their maxima are with larger magnitude than
those in the case with $c_1=0.60$ and in SuSA and their slopes are
steeper at large $T_{p(n)}$.

Secondly, the use of the exponential form of
$f_\text{RFG,2}^{p(n)}(\psi^\prime(R))$ (cases (b) and (d) in
Figs.~\ref{fig2ant}--\ref{fig5ant}) leads to a sharper slope of the
cross sections in comparison with the case of parabolic form (cases
(a) and (c)) for $T_{p(n)}$ smaller than those in the maximum.

Third, it can be seen from Figs.~\ref{fig2ant}--\ref{fig5ant} that,
similarly to the results in~\cite{ant26}, the shapes of the cross
sections for proton and neutron knockout are very similar. However,
the magnitudes are somewhat different. For instance, the magnitude
of the maximum of the NC cross section of neutrino scattering is
much larger than that for antineutrino scattering cross section.
This difference is around an order of magnitude in the case of
proton knockout at $\theta_p=20^\circ$, and it is even larger in the
case of neutron knockout at $\theta_n=20^\circ$.
%For the case of the proton knockout and $\theta_p=20^\circ$ this
%difference is around an order of magnitude and it is even more in
%the case of the neutron knockout.
For $\theta_p$ and $\theta_n$ equal to $60^\circ $ these differences
are smaller, around a factor of $2$.

Fourth, except for antineutrinos at forward angles, the neutron
knockout results are 30--50\% higher than the proton knockout. As
noted in~\cite{ant26}, this occurs because (in absence of
strangeness) both the vector and the axial-vector contributions are
larger for neutrons than for protons, and they sum up.

\section[]{CONCLUSIONS\label{sect4ant}}

In our previous work~\cite{ant19} we applied the superscaling
analysis and scaling functions obtained within the CDFM and LFD
approach to inclusive electron scattering as well as to
charge-changing neutrino and antineutrino reactions at energies
between 1 and 2~GeV from the $^{12}$C nucleus. The scaling functions
describe well the superscaling phenomenon below the QE peak.
In~\cite{ant19} the scaling function for the $\Delta$-region was
constructed and a good representation of inclusive electron
scattering cross sections data up to at least the peak of the
$\Delta $-region was obtained. The required asymmetry (with a long
tail extending to high energy loss) of the CDFM scaling function was
introduced in a phenomenological way.

%In the present work the superscaling  formalism was extended to
%include QE scattering via the weak neutral current of neutrinos and
%antineutrinos from nuclei at intermediate-to-high energies.
%Two forms of the asymmetric QE scaling function obtained in the CDFM
%were used in the present work to calculate the NC cross sections.
%One of them is with a parabolic form of the function
%$f_\text{RFG,2}^{p(n)}(\psi^\prime(R))$ [Eq.~(\ref{new24})] for
%positive values of $\psi^\prime$ used in Eqs.~(\ref{new10}),
%(\ref{new11}), (\ref{new14}) and it is the same as that employed in
%the CC studies. The second one is with an exponential form of the
%function $f_\text{RFG,2}^{p(n)}(\psi^\prime(R))$ [Eq.~(\ref{new15})]
%used in~(\ref{new10}), (\ref{new11}) and~(\ref{new14}).

In the present work we extend the application of CDFM scaling
functions to calculate differential cross sections of neutral
current neutrino- (antineutrino-) nucleus scattering at
intermediate-to-high energies. We construct asymmetric scaling
functions within CDFM taking into account the deviation from
experiment of the RFG scaling function at $\psi^\prime=0$ (see
Eqs.~(\ref{new10}) to (\ref{new14})). Two different asymmetric CDFM
scaling functions have been used in the present calculations. One
uses a parabolic form of the function
$f_\text{RFG,2}^{p(n)}(\psi^\prime(R))$ [Eq.~(\ref{new24})] at
positive $\psi^\prime$ values, as in our previous studies on charge
current neutrino scattering. The second uses an exponential form of
that function at $\psi^\prime>0$ [Eq.~(\ref{new15})].

In the CC studies the reaction involves an incoming lepton ($\nu$ or
$\bar{\nu}$) and the corresponding charged lepton is detected at a
given angle, just as in the case of the electron scattering with
incident and scattered electrons (both are $t$-channel inclusive
processes). In the NC reaction, in contrast to the CC process, one
has an incident $\nu$ or $\bar{\nu}$, but now a proton or neutron is
detected at some angle, the scattered $\nu $ or $\bar{\nu}$ not
being detected (this is the $u$-channel inclusive process). In this
work we adopt the $u$- versus $t$-channel scaling criteria
of~\cite{ant26} to apply the CDFM scaling functions to $u$-channel
scattering at intermediate-to-high energies.

It can be seen from our results at $60^\circ$ that the neutrino and
antineutrino cross sections are roughly in a $2:1$ ratio. For larger
scattering angle values, neutrino and antineutrino cross sections
come closer, diminishing the above ratio.
%It can be seen from our
%results that the neutrino and antineutrino cross sections are
%roughly in a $2:1$ ratio at $\nu N$ scattering angles larger than
%$60^\circ$.
At forward scattering angles the $\bar{\nu}$ cross
sections are strongly suppressed (by an order of magnitude or more).
This is observed for both proton and neutron knockout. Moreover, the
neutron knockout cross sections are somewhat larger than the proton
knockout cross sections due to the behavior of the NC single-nucleon
form factors.

It was shown that the use of asymmetric CDFM scaling function gives
results which are close to those from SuSA, while the symmetric
scaling function leads to a similarity with the RFG model results.
The asymmetric scaling function with an exponential form (by using
Eq.~(\ref{new15})) leads to a sharper slope of the cross sections,
in comparison to that with the parabolic form (by using
Eq.~(\ref{new24})), for the values of the kinetic energy $T_{p(n)}$
of the knocked-out nucleon smaller than those in the maximum of the
cross section.

In summary, we applied the superscaling approach by means of the
scaling function obtained within the CDFM (and used
previously~\cite{ant19} for the electron and CC neutrino reactions)
to the NC neutrino (antineutrino) scattering in the QE region at
energy of 1~GeV from the $^{12}$C nucleus. It is pointed out that
the constructed realistic CDFM scaling function is an essential
ingredient in this approach for the description of the processes of
lepton scattering from nuclei. Further, the CDFM model may also be
useful to explore to what extent the $u$- versus $t$-channel scaling
criteria, proposed in~\cite{ant26} on the basis of the RFG model,
may be proved to hold more generally.

Another interesting future project will be to extend the scaling
approach using a constructed realistic CDFM scaling function to
obtain predictions for charge-changing neutrino and antineutrino
scattering from nuclei in the $\Delta$-region.

\begin{acknowledgments}

One of the authors (M.K.G.) is grateful for the warm hospitality
given by the CSIC and for support during his stay there from the
State Secretariat of Education and Universities of Spain (N/Ref.
SAB2005--0012). This work was partly supported by the Bulgarian
National Science Fund under Contracts No.~$\Phi$--1416 and
$\Phi$--1501, and by Ministerio de Educaci\'on y Ciencia (Spain)
under contracts Nos.~FPA2006--13807--C02--01, FIS2005--01105, and
FIS2005--00640.

\end{acknowledgments}

\end{document}